\documentclass[pra,10pt,showpacs,eqsecnum,amsmath,twocolumn,
tightenlines,floatfix,notitlepage]{revtex4-1}
\usepackage[dvipsnames]{xcolor}
\usepackage[colorlinks=true,linkcolor=NavyBlue,urlcolor=NavyBlue,citecolor=NavyBlue,breaklinks]{hyperref}
\usepackage{amsmath}
\usepackage{amsfonts}
\usepackage{graphicx}
\usepackage{mathtools}

\newcommand{\intall}{\int_{-\infty}^{\infty}}
\newcommand{\ket}[1]{|#1\rangle}

\newcommand{\bra}[1]{\langle#1|}

\newcommand{\avg}[1]{\langle#1\rangle}
\newcommand{\Avg}[1]{\left\langle#1\right\rangle}

\newcommand{\erfc}{\operatorname{erfc}}

\newcommand{\abs}[1]{\left|#1\right|}
\newcommand{\bk}[1]{\left(#1\right)}
\newcommand{\Bk}[1]{\left[#1\right]}

\newcommand{\trace}{\operatorname{tr}}

\begin{document}
\title{Multiparameter Heisenberg limit}

\author{Mankei Tsang}

\email{eletmk@nus.edu.sg}
\affiliation{Department of Electrical and Computer Engineering,
  National University of Singapore, 4 Engineering Drive 3, Singapore
  117583}

\affiliation{Department of Physics, National University of Singapore,
  2 Science Drive 3, Singapore 117551}

\date{\today}

\begin{abstract}
  Using a quantum version of the Bell-Ziv-Zakai bound, I derive a
  Heisenberg limit to multiparameter estimation for any Gaussian prior
  probability density.  The mean-square error lower bound is shown to
  have a universal quadratic scaling with respect to a quantum resource,
  such as the average photon number in the case of optical phase estimation,
  suitably weighted by the prior covariance matrix. 
\end{abstract}

\maketitle
\section{Introduction}
The probabilistic nature of quantum mechanics imposes fundamental
limits to information processing applications
\cite{helstrom,glm_science,glm2011,holevo12}. Such quantum limits have
practical implications to many metrological applications, such as
optical interferometry, optomechanical sensing, gravitational-wave
detection \cite{braginsky,twc,tsang_nair,tsang_open}, optical imaging
\cite{treps,centroid,taylor2013}, magnetometry, gyroscopy, and atomic
clocks \cite{bollinger}. The existence of the so-called Heisenberg (H)
limit to parameter estimation has in particular attracted much
attention in recent years, as it implies that a minimum amount of
resource, such as the average photon number for optical phase
estimation, is needed to achieve a desired precision. After much
debate and confusion
\cite{yurke,sanders,ou,bollinger,zwierz,zwierz_err,rivas,luis_rodil,luis13,anisimov,zhang13},
it has now been proven that the H limit indeed exists for the
mean-square error of single-parameter estimation
\cite{qzzb,glm2012,hall2012,nair2012,gm_useless}. Although decoherence
can impose stricter limitations
\cite{knysh,escher,escher_bjp,escher_prl,latune,demkowicz,tsang_open,knysh14}
than the H limit, the latter can still be relevant when the
decoherence is relatively weak.

For many applications, such as waveform estimation
\cite{twc,tsang_open,bhw2013} and optical imaging \cite{humphreys}, the
estimation of multiple parameters from measurements is needed
\cite{yuen_lax,helstrom_kennedy,genoni}. In that case, the existence
of a general H limit remains an open question. A recent work by Zhang
and Fan \cite{zhang14} studies the quantum Ziv-Zakai bound (QZZB)
\cite{qzzb} for multiple parameters, but they assume that the
parameters are \textit{a priori} independent, such that the
single-parameter bound is applicable to each. In practice, and
especially for the waveform estimation problem, the parameters often
have nontrivial prior correlations, in which case a proper definition
of the relevant quantum resource is unknown and the H limit remains to
be proven.

Here I prove a multiparameter version of the H limit for any Gaussian
prior. The proof uses the Bell-Ziv-Zakai bound (BZZB) \cite{*[] [{,
    and references therein.}]  bell,bell1997}, which is an extension
of the Ziv-Zakai family of bounds for single-parameter estimation
\cite{bell}. The H limit is found to obey a universal quadratic
scaling with respect to a quantum resource suitably weighted by the
prior covariance matrix.  To illustrate the result, the bound is
applied to the problem of optical phase waveform estimation, showing
that an H limit can be defined with respect to the average photon
number within the prior correlation time scale of the waveform.

\section{Quantum Bell-Ziv-Zakai bound}
\begin{figure}[htbp]
\centerline{\includegraphics[width=0.45\textwidth]{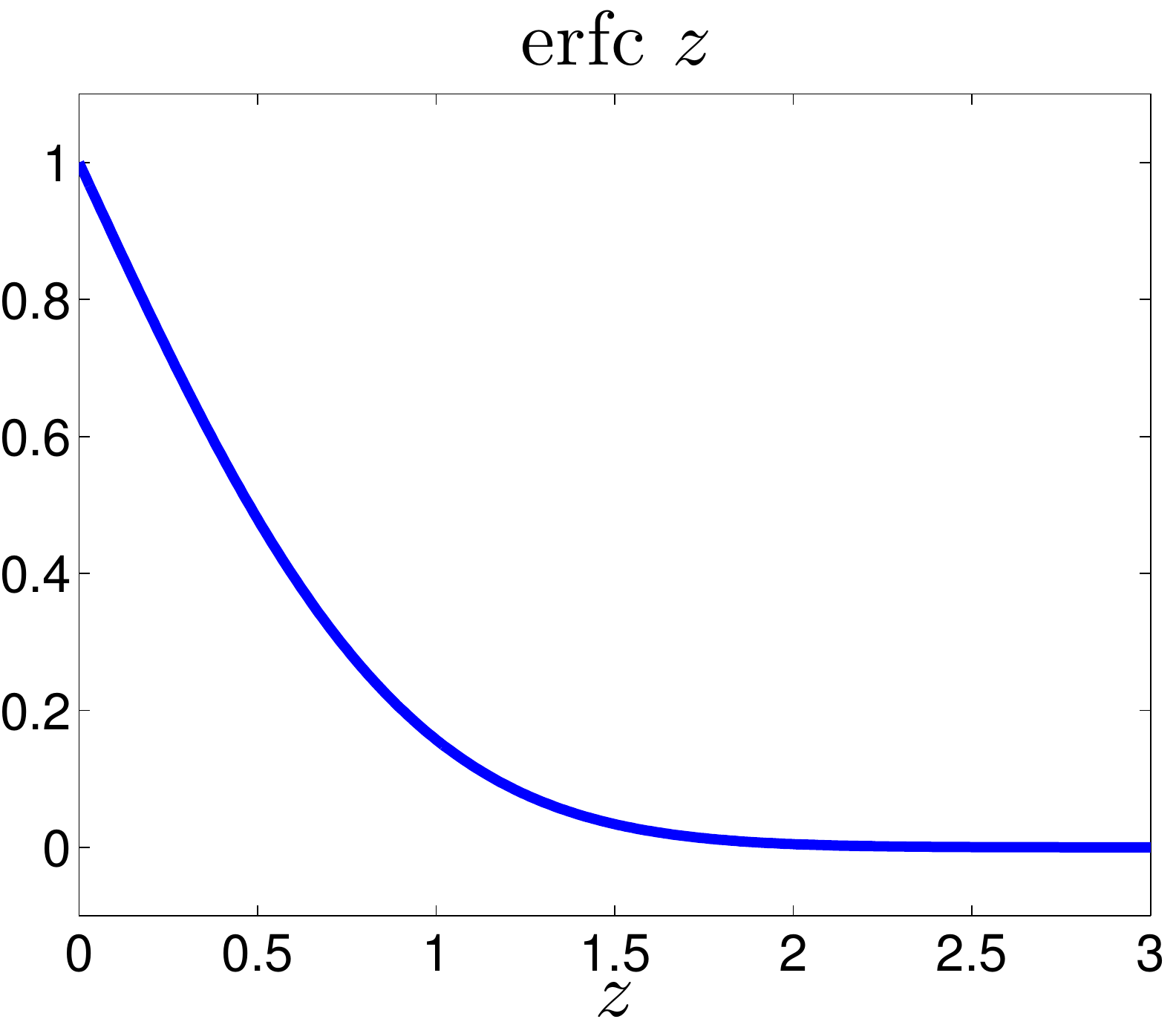}}
\caption{The erfc function.}
\label{erfc}
\end{figure}

Let $x$ be a column vector of the unknown parameters, $P(x)$ be its
prior probability density, $P(y|x)$ be the likelihood function with
observation $y$, and $\tilde x(y)$ be the estimator. The mean-square
error covariance matrix is defined as \cite{vantrees}
\begin{align}
\Sigma &\equiv \int dx dy P(y|x)P(x)\Bk{\tilde x(y)-x}\Bk{\tilde x(y)-x}^\top,
\end{align}
where $^\top$ denotes the transpose. One useful version of the BZZB is given by
\cite{bell,bell1997}
\begin{align}
u^\top \Sigma u &\ge \int_0^\infty d\tau \tau
\max_{v: u^\top v = 1}
\int dx \min\Bk{P(x), P(x+v \tau)}
\nonumber\\&\quad\times
P_e(x,x+v\tau),
\label{zzb}
\end{align}
where $u$ is an arbitrary real vector and $P_e(x^{(0)},x^{(1)})$ is
the error probability in discriminating equally likely hypotheses $x =
x^{(0)}$ and $x = x^{(1)}$ from an observation $y$ with the likelihood
function $P(y|x)$.  If $P_e(x,x+v\tau)$ does not depend on $x$, the
$x$ integral in Eq.~(\ref{zzb}) depends only on the prior distribution
$P(x)$. For a Gaussian $P(x)$ with covariance matrix $\Sigma_0$
\cite{bell1997},
\begin{align}
\int dx \min\Bk{P(x), P(x+v \tau)} = 
\erfc \frac{\tau}{\tau_0},
\nonumber\\
\erfc z \equiv \frac{2}{\sqrt\pi} \int_0^z d\xi \exp(-\xi^2),
\nonumber\\
\tau_0 \equiv \bk{\frac{8}{v^\top \Sigma_0^{-1} v}}^{1/2}.
\label{gauss}
\end{align}
The erfc function is plotted in Fig.~\ref{erfc}.

Suppose now that a quantum probe is used to measure the parameters.
The likelihood function becomes
\begin{align}
P(y|x) &= \trace E(y) \rho_x,
\end{align}
where $E(y)$ is the positive operator-valued measure (POVM) that
describes the measurement and $\rho_x$ is the density operator
conditioned on the unknown $x$.  The following quantum bound can be
used \cite{fuchs}:
\begin{align}
P_e(x,x+v\tau) &\ge \frac{1}{2}\Bk{1-\sqrt{1-F(\rho_x,\rho_{x+v\tau})}},
\label{helstrom}
\end{align}
where
\begin{align}
F(\rho_x,\rho_{x+v\tau}) &\equiv 
\bk{\trace \sqrt{\sqrt{\rho_x} \rho_{x+v\tau}\sqrt{\rho_x}}}^2
\end{align}
is the Uhlmann fidelity between $\rho_x$ and $\rho_{x+v\tau}$.  This
quantum bound, together with the BZZB, results in a quantum
Bell-Ziv-Zakai bound (QBZZB) on the mean-square error of
multiparameter estimation, just like the single-parameter case
\cite{qzzb}.  It is possible to define QBZZBs for error functions
other than the mean-square criterion \cite{bell,bell1997}, although I
shall focus on the mean-square error here because of its popularity.

\section{Quantum phase estimation}
Suppose that the density operator is
\begin{align}
\rho_x &= U_x \rho U_x^\dagger,
\end{align}
and the unitary has the following form:
\begin{align}
U_x &= \exp(i x^\top n) = \exp\bigg(i\sum_j x_j n_j \bigg),
\end{align}
where $n$ is a column vector of quantum operators and $\rho$ is the
initial density operator. Assuming that $\ket\psi$ is a purification
of $\rho$ and defining
\begin{align}
\avg{O} \equiv \bra{\psi}O\ket{\psi},
\end{align}
a lower bound on the fidelity is given by
\begin{align}
F(\rho_x,\rho_{x+v\tau}) &\ge \abs{\Avg{\exp(i\tau v^\top n)}}^2
\\
&= \sum_{m,l} P_m P_l \exp[i\tau v^\top (m-l)]
\\
&= \sum_{m,l} P_m P_l \cos[\tau v^\top (m-l)],
\label{cosine}
\end{align}
where 
\begin{align}
P_m &\equiv \abs{\avg{m|\psi}}^2
\end{align}
is the probability distribution with respect to the $n$ eigenstates.

\begin{figure}[htbp]
\centerline{\includegraphics[width=0.45\textwidth]{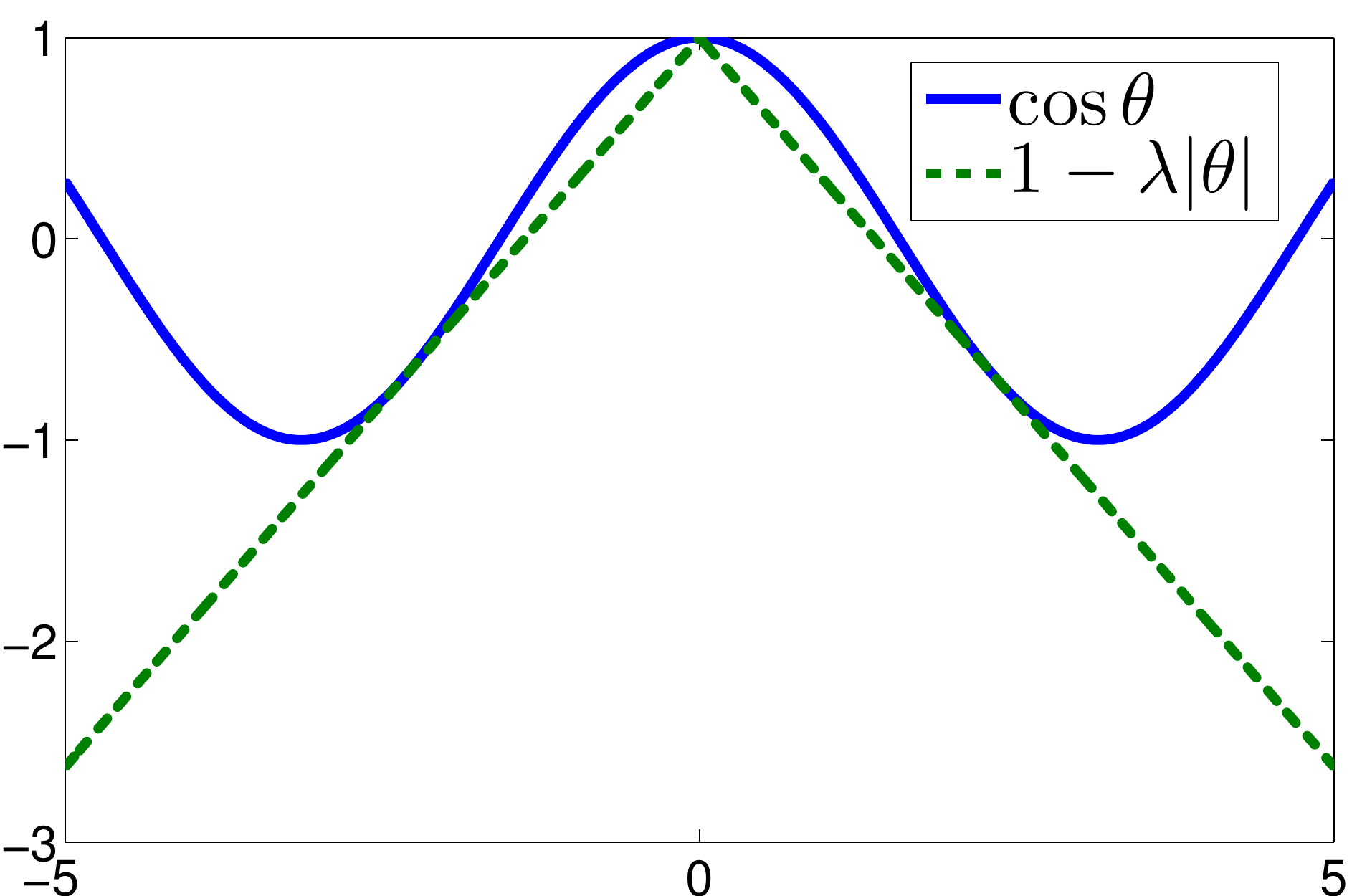}}
\caption{A lower bound for cosine.}
\label{cosine_bound}
\end{figure}

\begin{figure}[htbp]
\centerline{\includegraphics[width=0.45\textwidth]{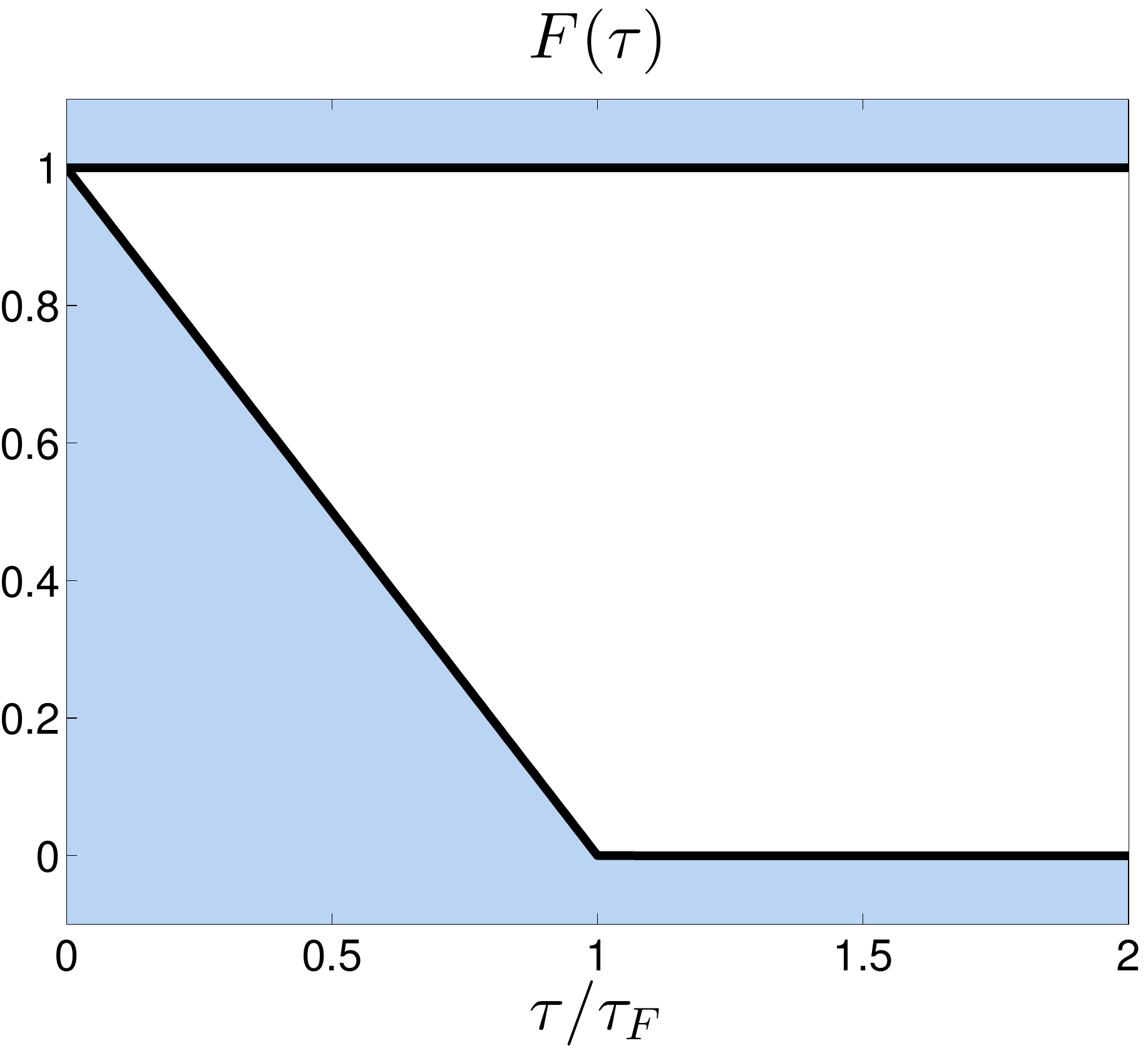}}
\caption{Bounds for the fidelity. The white area is the permissable area.}
\label{triangle}
\end{figure}

A useful bound for the cosine function for deriving the H
limit is \cite{qzzb}
\begin{align}
\cos \theta &\ge 1-\lambda|\theta|,
\label{bound}
\end{align}
where $\lambda \approx 0.7246$ is a solution of $\lambda = \sin\phi =
(1-\cos\phi)/\phi$, as shown in Fig.~\ref{cosine_bound}.  Substituting
this bound into Eq.~(\ref{cosine}) and using the triangle inequality,
one obtains
\begin{align}
F &\ge 
\sum_{m,l} P_m P_l\Bk{1-\lambda \tau|v^\top (m-l)|}
\\
&\ge \sum_{m,l} P_m P_l\Bk{1-\lambda \tau\bk{|v^\top m - H_0|+|v^\top l-H_0|}}
\\
&= 1- 2\lambda \tau \avg{|v^\top n-H_0|},
\label{speed}
\end{align}
where $H_0$ is an arbitrary constant. It is possible to obtain a
slightly tighter bound numerically using the method in
Refs.~\cite{glm_speed,gm_useless}, but Eq.~(\ref{speed}) will produce
the same scaling.  Since $0\le F\le 1$, a tighter lower
bound is
\begin{align}
F &\ge \Lambda\bk{\frac{\tau}{\tau_F}}\equiv \Big\{
\begin{array}{ll}
1- \tau/\tau_F, & \tau < \tau_F,
\\
0, & \tau \ge \tau_F,
\end{array}
\nonumber\\
\tau_F &\equiv \frac{1}{2\lambda\avg{|v^\top n-H_0|}},
\label{F_bound}
\end{align}
as shown in Fig.~\ref{triangle}.

\begin{figure}[htbp]
\centerline{\includegraphics[width=0.45\textwidth]{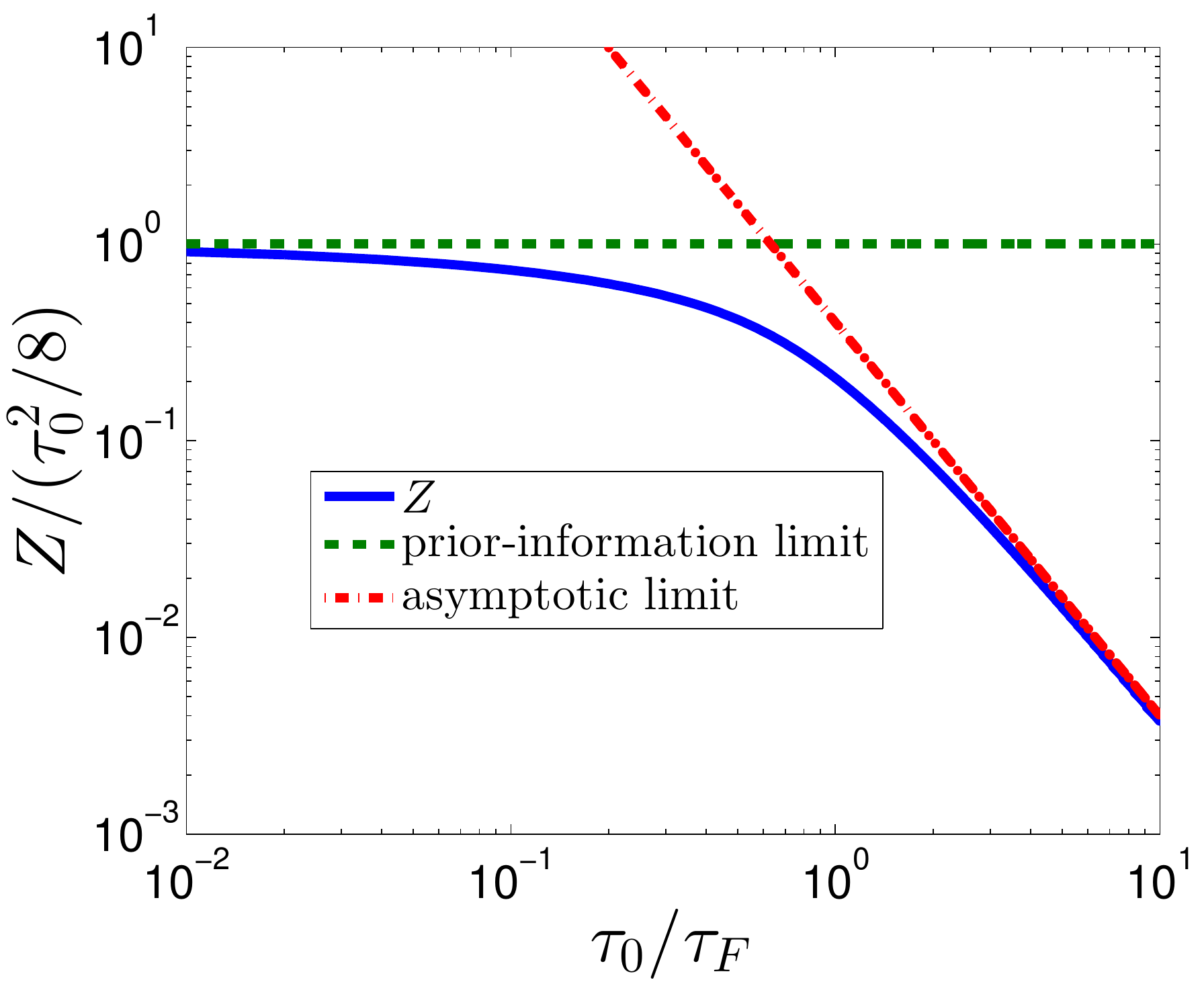}}
\caption{A quantum lower error bound $Z$ on $u^\top\Sigma u$ versus
  the parameter $\tau_0/\tau_F$ in log-log scale, the
  prior-information limit $Z \to \tau_0^2/8$, and the asymptotic H
  limit $Z \to \tau_F^2/20$.}
\label{qbzzb}
\end{figure}

Putting Eqs.~(\ref{gauss}), (\ref{helstrom}), and (\ref{F_bound}) together,
\begin{align}
&\quad\max_{v: u^\top v = 1}
\int dx \min\Bk{P(x), P(x+v \tau)}
P_e(x,x+v\tau) 
\nonumber\\
&\ge
\frac{1}{2}\max_{v: u^\top v = 1} 
\erfc\bk{\frac{\tau}{\tau_0}}\Lambda\bk{\sqrt{\frac{\tau}{\tau_F}}}.
\label{max_v}
\end{align}
Recall that $\tau_0$ and $\tau_F$ depend on $v$. The maximization does
not seem to be tractable analytically, so I choose a $v$ that
maximizes only the erfc function:
\begin{align}
v_0 &\equiv \arg \max_{v: u^\top v = 1} 
\erfc\bk{\frac{\tau}{\tau_0}} = \frac{\Sigma_0 u}{u^\top \Sigma_0 u},
\end{align}
such that 
\begin{align}
&\quad \max_{v: u^\top v = 1} 
\erfc\bk{\frac{\tau}{\tau_0}}\Lambda\bk{\sqrt{\frac{\tau}{\tau_F}}}
\nonumber\\
&\ge 
\erfc\bk{\frac{\tau}{\tau_0}}\Lambda\bk{\sqrt{\frac{\tau}{\tau_F}}}\bigg|_{v = v_0},
\label{v0}
\nonumber\\
\tau_0(v_0) &= 2\sqrt{2 u^\top\Sigma_0 u},
\nonumber\\
\tau_F(v_0) &= \frac{1}{2\lambda\avg{|u^\top\Sigma_0 n/(u^\top\Sigma_0 u) - H_0|}}.
\end{align}
Combining Eqs.~(\ref{zzb}), (\ref{max_v}), and (\ref{v0}) then
produces the following bound:
\begin{align}
u^\top\Sigma u &\ge  Z\equiv \frac{1}{2}\int_0^{\tau_F} d\tau \tau 
\erfc\bk{\frac{\tau}{\tau_0}}\bk{1-\sqrt{\frac{\tau}{\tau_F}}}\bigg|_{v = v_0}
\label{Z}
\end{align}
The integral can be computed numerically, as shown in Fig.~\ref{qbzzb},
but there are two analytic limits of interest:
\begin{enumerate}
\item The prior-information limit ($\tau_F\gg \tau_0$):
\begin{align}
\lim_{\tau_F/\tau_0\to \infty} Z &= \frac{\tau_0^2}{8} = u^\top \Sigma_0 u,
\end{align}
where the bound is determined only by the prior covariance matrix, as
expected;
\item The asymptotic limit ($\tau_F \ll \tau_0$), where the
  measurement provides much more information:
\begin{align}
\lim_{\tau_0/\tau_F\to \infty} Z &= \frac{\tau_F^2}{20} = 
\frac{1}{80\lambda^2 H_+^2},
\nonumber\\
H_+ &\equiv  \Avg{\abs{\frac{u^\top\Sigma_0 n}{u^\top\Sigma_0 u}-H_0}},
\label{central}
\end{align}
and $H_+$ quantifies the relevant resource for the estimation.
Eq.~(\ref{central}) is the central result of this paper and an
appropriate generalization of the single-parameter case \cite{qzzb}.
\end{enumerate}

For example, the error bound for estimating a particular parameter
$x_k$ can be obtained by setting $u$ as
\begin{align}
u_j &= \delta_{jk},
\\
u^\top\Sigma u &= \Sigma_{kk} \ge Z_k \to 
\frac{1}{80\lambda^2 H_{+k}^2},
\\
H_{+k} &\equiv 
\Avg{\abs{\frac{1}{\Sigma_{0kk}}\sum_l \Sigma_{0 kl} n_l-H_0}}.
\label{hlimit}
\end{align}
For optical phase estimation with $n_l$ being a photon number
operator, one can assume $H_0 = 0$ and use the triangle inequality to
obtain
\begin{align}
H_{+k} &\le 
\frac{1}{\Sigma_{0kk}}\sum_l |\Sigma_{0 kl}| \Avg{n_l},
\end{align}
which produces an H limit with respect to a weighted average of the
photon numbers. The weighting of the photon numbers with respect to
the prior covariance matrix is the key feature of the bound, as it
properly accounts for the optical modes that can contribute to the
estimation of a particular phase.

A special case is when the parameters are independent \textit{a
  priori}, such that
\begin{align}
\Sigma_{0kl} &= \Sigma_{0kk}\delta_{kl},
\\
H_{+k} &= \avg{|n_k-H_0|},
\end{align}
and the single-parameter bound \cite{qzzb} is recovered. Zhang and Fan
used this \cite{zhang14} to rule out any significant quantum
enhancement with a proposal by Humphreys \textit{et al.}\ for quantum
multiparameter estimation \cite{humphreys}.

\section{Optical phase waveform estimation}
To illustrate the result derived in the previous section, consider the
continuous-time limit of the QBZZB for optical phase estimation.  The
photon number of each mode is related to the photon flux $I(t)$ and
the time duration $dt$ of the mode:
\begin{align}
n_l &= dt I(t_l).
\end{align}
The mean-square error for each phase parameter becomes
the error for estimating the phase at a certain time:
\begin{align}
\Sigma_{kk} &= \Sigma(t_k,t_k),
\end{align}
and the H limit becomes
\begin{align}
\Sigma(t,t) &\ge Z(t)\to  \frac{1}{80\lambda^2 H_+^2(t)},
\\
H_+(t) &\equiv
\Avg{\abs{\frac{1}{\Sigma_0(t,t)}\int dt'  \Sigma_0(t,t') I(t')-H_0}}
\\
&\le  \frac{1}{\Sigma_0(t,t)}\int dt' |\Sigma_0(t,t')| \Avg{I(t')}.
\end{align}
The relevant resource $H_+$ is defined as the time integral of the
average photon flux $\avg{I(t')}$ weighted by the prior covariance
function $\Sigma_0(t,t')$.  For example, for the Ornstein-Uhlenbeck
process,
\begin{align}
\Sigma_0(t,t') &= \sigma_0\exp\bk{-\frac{|t-t'|}{T_0}},
\\
H_+(t) &\le  \intall dt' \exp\bk{-\frac{|t-t'|}{T_0}} \Avg{I(t')},
\end{align}
which states that only the optical modes within the prior time scale
$T_0$ can contribute to the estimation at a particular time.

If $\avg{I}$ is constant in time, $H_+(t)\propto \avg{I}$, and there
exists a universal quadratic error scaling $\propto 1/\avg{I}^2$ for
any Gaussian prior.  Tighter scalings can be derived for Gaussian
quantum states \cite{bhw2013}, but the H limit is still valuable as a
simple and more general no-go theorem.

\section{Conclusion}
To conclude, I have proved an H limit with a universal $1/N^2$ scaling
for multiparameter estimation with any Gaussian prior, where $N$ is an
appropriately defined quantum resource. The key feature of the bound
is the use of the prior covariance matrix to define $N$, enabling a
proper accounting of the relevant quantum resources. In the case of
optical phase waveform estimation, the H limit implies the intuitive
result that only the optical modes within the prior correlation time
scale can contribute to the estimation at a particular time.

It should be emphasized that the H limit derived here may well not be
attainable and the quantum Cram\'er-Rao bound
\cite{twc,tsang_open,bhw2013} may provide tighter bounds for more
specific quantum states, but the generality and simplicity of the
result here should still be valuable as a no-go theorem. It may also
be possible to derive tighter bounds or study other priors using the
present formalism. These possibilities are left for future
investigations.

\section*{Acknowledgments}
Discussions with Ranjith Nair are gratefully acknowledged.
This work is supported by the Singapore National Research
Foundation under NRF Grant No. NRF-NRFF2011-07.

\bibliography{research}

\begin{thebibliography}{45}%
\makeatletter
\providecommand \@ifxundefined [1]{%
 \@ifx{#1\undefined}
}%
\providecommand \@ifnum [1]{%
 \ifnum #1\expandafter \@firstoftwo
 \else \expandafter \@secondoftwo
 \fi
}%
\providecommand \@ifx [1]{%
 \ifx #1\expandafter \@firstoftwo
 \else \expandafter \@secondoftwo
 \fi
}%
\providecommand \natexlab [1]{#1}%
\providecommand \enquote  [1]{``#1''}%
\providecommand \bibnamefont  [1]{#1}%
\providecommand \bibfnamefont [1]{#1}%
\providecommand \citenamefont [1]{#1}%
\providecommand \href@noop [0]{\@secondoftwo}%
\providecommand \href [0]{\begingroup \@sanitize@url \@href}%
\providecommand \@href[1]{\@@startlink{#1}\@@href}%
\providecommand \@@href[1]{\endgroup#1\@@endlink}%
\providecommand \@sanitize@url [0]{\catcode `\\12\catcode `\$12\catcode
  `\&12\catcode `\#12\catcode `\^12\catcode `\_12\catcode `\%12\relax}%
\providecommand \@@startlink[1]{}%
\providecommand \@@endlink[0]{}%
\providecommand \url  [0]{\begingroup\@sanitize@url \@url }%
\providecommand \@url [1]{\endgroup\@href {#1}{\urlprefix }}%
\providecommand \urlprefix  [0]{URL }%
\providecommand \Eprint [0]{\href }%
\providecommand \doibase [0]{http://dx.doi.org/}%
\providecommand \selectlanguage [0]{\@gobble}%
\providecommand \bibinfo  [0]{\@secondoftwo}%
\providecommand \bibfield  [0]{\@secondoftwo}%
\providecommand \translation [1]{[#1]}%
\providecommand \BibitemOpen [0]{}%
\providecommand \bibitemStop [0]{}%
\providecommand \bibitemNoStop [0]{.\EOS\space}%
\providecommand \EOS [0]{\spacefactor3000\relax}%
\providecommand \BibitemShut  [1]{\csname bibitem#1\endcsname}%
\let\auto@bib@innerbib\@empty
\bibitem [{\citenamefont {Helstrom}(1976)}]{helstrom}%
  \BibitemOpen
  \bibfield  {author} {\bibinfo {author} {\bibfnamefont {C.~W.}\ \bibnamefont
  {Helstrom}},\ }\href@noop {} {\emph {\bibinfo {title} {Quantum Detection and
  Estimation Theory}}}\ (\bibinfo  {publisher} {Academic Press},\ \bibinfo
  {address} {New York},\ \bibinfo {year} {1976})\BibitemShut {NoStop}%
\bibitem [{\citenamefont {Giovannetti}\ \emph {et~al.}(2004)\citenamefont
  {Giovannetti}, \citenamefont {Lloyd},\ and\ \citenamefont
  {Maccone}}]{glm_science}%
  \BibitemOpen
  \bibfield  {author} {\bibinfo {author} {\bibfnamefont {V.}~\bibnamefont
  {Giovannetti}}, \bibinfo {author} {\bibfnamefont {S.}~\bibnamefont {Lloyd}},
  \ and\ \bibinfo {author} {\bibfnamefont {L.}~\bibnamefont {Maccone}},\ }\href
  {\doibase 10.1126/science.1104149} {\bibfield  {journal} {\bibinfo  {journal}
  {Science}\ }\textbf {\bibinfo {volume} {306}},\ \bibinfo {pages} {1330}
  (\bibinfo {year} {2004})}\BibitemShut {NoStop}%
\bibitem [{\citenamefont {Giovannetti}\ \emph {et~al.}(2011)\citenamefont
  {Giovannetti}, \citenamefont {Lloyd},\ and\ \citenamefont
  {Maccone}}]{glm2011}%
  \BibitemOpen
  \bibfield  {author} {\bibinfo {author} {\bibfnamefont {V.}~\bibnamefont
  {Giovannetti}}, \bibinfo {author} {\bibfnamefont {S.}~\bibnamefont {Lloyd}},
  \ and\ \bibinfo {author} {\bibfnamefont {L.}~\bibnamefont {Maccone}},\ }\href
  {http://dx.doi.org/10.1038/nphoton.2011.35} {\bibfield  {journal} {\bibinfo
  {journal} {Nature Photon.}\ }\textbf {\bibinfo {volume} {5}},\ \bibinfo
  {pages} {222} (\bibinfo {year} {2011})}\BibitemShut {NoStop}%
\bibitem [{\citenamefont {Holevo}(2012)}]{holevo12}%
  \BibitemOpen
  \bibfield  {author} {\bibinfo {author} {\bibfnamefont {A.~S.}\ \bibnamefont
  {Holevo}},\ }\href {http://www.degruyter.com/view/product/180320} {\emph
  {\bibinfo {title} {Quantum Systems, Channels, Information}}}\ (\bibinfo
  {publisher} {de Gruyter},\ \bibinfo {address} {Berlin},\ \bibinfo {year}
  {2012})\BibitemShut {NoStop}%
\bibitem [{\citenamefont {Braginsky}\ and\ \citenamefont
  {Khalili}(1992)}]{braginsky}%
  \BibitemOpen
  \bibfield  {author} {\bibinfo {author} {\bibfnamefont {V.~B.}\ \bibnamefont
  {Braginsky}}\ and\ \bibinfo {author} {\bibfnamefont {F.~Y.}\ \bibnamefont
  {Khalili}},\ }\href@noop {} {\emph {\bibinfo {title} {Quantum Measurement}}}\
  (\bibinfo  {publisher} {Cambridge University Press},\ \bibinfo {address}
  {Cambridge},\ \bibinfo {year} {1992})\BibitemShut {NoStop}%
\bibitem [{\citenamefont {Tsang}\ \emph {et~al.}(2011)\citenamefont {Tsang},
  \citenamefont {Wiseman},\ and\ \citenamefont {Caves}}]{twc}%
  \BibitemOpen
  \bibfield  {author} {\bibinfo {author} {\bibfnamefont {M.}~\bibnamefont
  {Tsang}}, \bibinfo {author} {\bibfnamefont {H.~M.}\ \bibnamefont {Wiseman}},
  \ and\ \bibinfo {author} {\bibfnamefont {C.~M.}\ \bibnamefont {Caves}},\
  }\href {\doibase 10.1103/PhysRevLett.106.090401} {\bibfield  {journal}
  {\bibinfo  {journal} {Phys. Rev. Lett.}\ }\textbf {\bibinfo {volume} {106}},\
  \bibinfo {pages} {090401} (\bibinfo {year} {2011})}\BibitemShut {NoStop}%
\bibitem [{\citenamefont {Tsang}\ and\ \citenamefont
  {Nair}(2012)}]{tsang_nair}%
  \BibitemOpen
  \bibfield  {author} {\bibinfo {author} {\bibfnamefont {M.}~\bibnamefont
  {Tsang}}\ and\ \bibinfo {author} {\bibfnamefont {R.}~\bibnamefont {Nair}},\
  }\href {\doibase 10.1103/PhysRevA.86.042115} {\bibfield  {journal} {\bibinfo
  {journal} {Phys. Rev. A}\ }\textbf {\bibinfo {volume} {86}},\ \bibinfo
  {pages} {042115} (\bibinfo {year} {2012})}\BibitemShut {NoStop}%
\bibitem [{\citenamefont {Tsang}(2013)}]{tsang_open}%
  \BibitemOpen
  \bibfield  {author} {\bibinfo {author} {\bibfnamefont {M.}~\bibnamefont
  {Tsang}},\ }\href {\doibase 10.1088/1367-2630/15/7/073005} {\bibfield
  {journal} {\bibinfo  {journal} {New Journal of Physics}\ }\textbf {\bibinfo
  {volume} {15}},\ \bibinfo {pages} {073005} (\bibinfo {year}
  {2013})}\BibitemShut {NoStop}%
\bibitem [{\citenamefont {Treps}\ \emph {et~al.}(2003)\citenamefont {Treps},
  \citenamefont {Grosse}, \citenamefont {Bowen}, \citenamefont {Fabre},
  \citenamefont {Bachor},\ and\ \citenamefont {Lam}}]{treps}%
  \BibitemOpen
  \bibfield  {author} {\bibinfo {author} {\bibfnamefont {N.}~\bibnamefont
  {Treps}}, \bibinfo {author} {\bibfnamefont {N.}~\bibnamefont {Grosse}},
  \bibinfo {author} {\bibfnamefont {W.~P.}\ \bibnamefont {Bowen}}, \bibinfo
  {author} {\bibfnamefont {C.}~\bibnamefont {Fabre}}, \bibinfo {author}
  {\bibfnamefont {H.-A.}\ \bibnamefont {Bachor}}, \ and\ \bibinfo {author}
  {\bibfnamefont {P.~K.}\ \bibnamefont {Lam}},\ }\href {\doibase
  10.1126/science.1086489} {\bibfield  {journal} {\bibinfo  {journal}
  {Science}\ }\textbf {\bibinfo {volume} {301}},\ \bibinfo {pages} {940}
  (\bibinfo {year} {2003})},\ \Eprint
  {http://arxiv.org/abs/http://www.sciencemag.org/content/301/5635/940.full.pdf}
  {http://www.sciencemag.org/content/301/5635/940.full.pdf} \BibitemShut
  {NoStop}%
\bibitem [{\citenamefont {Tsang}(2009)}]{centroid}%
  \BibitemOpen
  \bibfield  {author} {\bibinfo {author} {\bibfnamefont {M.}~\bibnamefont
  {Tsang}},\ }\href {\doibase 10.1103/PhysRevLett.102.253601} {\bibfield
  {journal} {\bibinfo  {journal} {Phys. Rev. Lett.}\ }\textbf {\bibinfo
  {volume} {102}},\ \bibinfo {pages} {253601} (\bibinfo {year}
  {2009})}\BibitemShut {NoStop}%
\bibitem [{\citenamefont {Taylor}\ \emph {et~al.}(2013)\citenamefont {Taylor},
  \citenamefont {Janousek}, \citenamefont {Daria}, \citenamefont {Knittel},
  \citenamefont {Hage}, \citenamefont {Bachor},\ and\ \citenamefont
  {Bowen}}]{taylor2013}%
  \BibitemOpen
  \bibfield  {author} {\bibinfo {author} {\bibfnamefont {M.~A.}\ \bibnamefont
  {Taylor}}, \bibinfo {author} {\bibfnamefont {J.}~\bibnamefont {Janousek}},
  \bibinfo {author} {\bibfnamefont {V.}~\bibnamefont {Daria}}, \bibinfo
  {author} {\bibfnamefont {J.}~\bibnamefont {Knittel}}, \bibinfo {author}
  {\bibfnamefont {B.}~\bibnamefont {Hage}}, \bibinfo {author} {\bibfnamefont
  {H.-A.}\ \bibnamefont {Bachor}}, \ and\ \bibinfo {author} {\bibfnamefont
  {W.~P.}\ \bibnamefont {Bowen}},\ }\href@noop {} {\bibfield  {journal}
  {\bibinfo  {journal} {Nature Photonics}\ }\textbf {\bibinfo {volume} {7}},\
  \bibinfo {pages} {229} (\bibinfo {year} {2013})}\BibitemShut {NoStop}%
\bibitem [{\citenamefont {Bollinger}\ \emph {et~al.}(1996)\citenamefont
  {Bollinger}, \citenamefont {Itano}, \citenamefont {Wineland},\ and\
  \citenamefont {Heinzen}}]{bollinger}%
  \BibitemOpen
  \bibfield  {author} {\bibinfo {author} {\bibfnamefont {J.~J.~.}\ \bibnamefont
  {Bollinger}}, \bibinfo {author} {\bibfnamefont {W.~M.}\ \bibnamefont
  {Itano}}, \bibinfo {author} {\bibfnamefont {D.~J.}\ \bibnamefont {Wineland}},
  \ and\ \bibinfo {author} {\bibfnamefont {D.~J.}\ \bibnamefont {Heinzen}},\
  }\href {\doibase 10.1103/PhysRevA.54.R4649} {\bibfield  {journal} {\bibinfo
  {journal} {Phys. Rev. A}\ }\textbf {\bibinfo {volume} {54}},\ \bibinfo
  {pages} {R4649} (\bibinfo {year} {1996})}\BibitemShut {NoStop}%
\bibitem [{\citenamefont {Yurke}\ \emph {et~al.}(1986)\citenamefont {Yurke},
  \citenamefont {McCall},\ and\ \citenamefont {Klauder}}]{yurke}%
  \BibitemOpen
  \bibfield  {author} {\bibinfo {author} {\bibfnamefont {B.}~\bibnamefont
  {Yurke}}, \bibinfo {author} {\bibfnamefont {S.~L.}\ \bibnamefont {McCall}}, \
  and\ \bibinfo {author} {\bibfnamefont {J.~R.}\ \bibnamefont {Klauder}},\
  }\href {\doibase 10.1103/PhysRevA.33.4033} {\bibfield  {journal} {\bibinfo
  {journal} {Phys. Rev. A}\ }\textbf {\bibinfo {volume} {33}},\ \bibinfo
  {pages} {4033} (\bibinfo {year} {1986})}\BibitemShut {NoStop}%
\bibitem [{\citenamefont {Sanders}\ and\ \citenamefont
  {Milburn}(1995)}]{sanders}%
  \BibitemOpen
  \bibfield  {author} {\bibinfo {author} {\bibfnamefont {B.~C.}\ \bibnamefont
  {Sanders}}\ and\ \bibinfo {author} {\bibfnamefont {G.~J.}\ \bibnamefont
  {Milburn}},\ }\href {\doibase 10.1103/PhysRevLett.75.2944} {\bibfield
  {journal} {\bibinfo  {journal} {Phys. Rev. Lett.}\ }\textbf {\bibinfo
  {volume} {75}},\ \bibinfo {pages} {2944} (\bibinfo {year}
  {1995})}\BibitemShut {NoStop}%
\bibitem [{\citenamefont {Ou}(1996)}]{ou}%
  \BibitemOpen
  \bibfield  {author} {\bibinfo {author} {\bibfnamefont {Z.~Y.}\ \bibnamefont
  {Ou}},\ }\href {\doibase 10.1103/PhysRevLett.77.2352} {\bibfield  {journal}
  {\bibinfo  {journal} {Phys. Rev. Lett.}\ }\textbf {\bibinfo {volume} {77}},\
  \bibinfo {pages} {2352} (\bibinfo {year} {1996})}\BibitemShut {NoStop}%
\bibitem [{\citenamefont {Zwierz}\ \emph {et~al.}(2010)\citenamefont {Zwierz},
  \citenamefont {P\'erez-Delgado},\ and\ \citenamefont {Kok}}]{zwierz}%
  \BibitemOpen
  \bibfield  {author} {\bibinfo {author} {\bibfnamefont {M.}~\bibnamefont
  {Zwierz}}, \bibinfo {author} {\bibfnamefont {C.~A.}\ \bibnamefont
  {P\'erez-Delgado}}, \ and\ \bibinfo {author} {\bibfnamefont {P.}~\bibnamefont
  {Kok}},\ }\href {\doibase 10.1103/PhysRevLett.105.180402} {\bibfield
  {journal} {\bibinfo  {journal} {Phys. Rev. Lett.}\ }\textbf {\bibinfo
  {volume} {105}},\ \bibinfo {pages} {180402} (\bibinfo {year}
  {2010})}\BibitemShut {NoStop}%
\bibitem [{\citenamefont {Zwierz}\ \emph {et~al.}(2011)\citenamefont {Zwierz},
  \citenamefont {P\'erez-Delgado},\ and\ \citenamefont {Kok}}]{zwierz_err}%
  \BibitemOpen
  \bibfield  {author} {\bibinfo {author} {\bibfnamefont {M.}~\bibnamefont
  {Zwierz}}, \bibinfo {author} {\bibfnamefont {C.~A.}\ \bibnamefont
  {P\'erez-Delgado}}, \ and\ \bibinfo {author} {\bibfnamefont {P.}~\bibnamefont
  {Kok}},\ }\href {\doibase 10.1103/PhysRevLett.107.059904} {\bibfield
  {journal} {\bibinfo  {journal} {Phys. Rev. Lett.}\ }\textbf {\bibinfo
  {volume} {107}},\ \bibinfo {pages} {059904} (\bibinfo {year}
  {2011})}\BibitemShut {NoStop}%
\bibitem [{\citenamefont {Rivas}\ and\ \citenamefont {Luis}(2012)}]{rivas}%
  \BibitemOpen
  \bibfield  {author} {\bibinfo {author} {\bibfnamefont {{\~{A}}.}~\bibnamefont
  {Rivas}}\ and\ \bibinfo {author} {\bibfnamefont {A.}~\bibnamefont {Luis}},\
  }\href {http://stacks.iop.org/1367-2630/14/i=9/a=093052} {\bibfield
  {journal} {\bibinfo  {journal} {New Journal of Physics}\ }\textbf {\bibinfo
  {volume} {14}},\ \bibinfo {pages} {093052} (\bibinfo {year}
  {2012})}\BibitemShut {NoStop}%
\bibitem [{\citenamefont {Luis}\ and\ \citenamefont
  {Rodil}(2013)}]{luis_rodil}%
  \BibitemOpen
  \bibfield  {author} {\bibinfo {author} {\bibfnamefont {A.}~\bibnamefont
  {Luis}}\ and\ \bibinfo {author} {\bibfnamefont {A.}~\bibnamefont {Rodil}},\
  }\href {\doibase 10.1103/PhysRevA.87.034101} {\bibfield  {journal} {\bibinfo
  {journal} {Phys. Rev. A}\ }\textbf {\bibinfo {volume} {87}},\ \bibinfo
  {pages} {034101} (\bibinfo {year} {2013})}\BibitemShut {NoStop}%
\bibitem [{\citenamefont {Luis}(2013)}]{luis13}%
  \BibitemOpen
  \bibfield  {author} {\bibinfo {author} {\bibfnamefont {A.}~\bibnamefont
  {Luis}},\ }\href {\doibase http://dx.doi.org/10.1016/j.aop.2012.12.004}
  {\bibfield  {journal} {\bibinfo  {journal} {Annals of Physics}\ }\textbf
  {\bibinfo {volume} {331}},\ \bibinfo {pages} {1 } (\bibinfo {year}
  {2013})}\BibitemShut {NoStop}%
\bibitem [{\citenamefont {Anisimov}\ \emph {et~al.}(2010)\citenamefont
  {Anisimov}, \citenamefont {Raterman}, \citenamefont {Chiruvelli},
  \citenamefont {Plick}, \citenamefont {Huver}, \citenamefont {Lee},\ and\
  \citenamefont {Dowling}}]{anisimov}%
  \BibitemOpen
  \bibfield  {author} {\bibinfo {author} {\bibfnamefont {P.~M.}\ \bibnamefont
  {Anisimov}}, \bibinfo {author} {\bibfnamefont {G.~M.}\ \bibnamefont
  {Raterman}}, \bibinfo {author} {\bibfnamefont {A.}~\bibnamefont
  {Chiruvelli}}, \bibinfo {author} {\bibfnamefont {W.~N.}\ \bibnamefont
  {Plick}}, \bibinfo {author} {\bibfnamefont {S.~D.}\ \bibnamefont {Huver}},
  \bibinfo {author} {\bibfnamefont {H.}~\bibnamefont {Lee}}, \ and\ \bibinfo
  {author} {\bibfnamefont {J.~P.}\ \bibnamefont {Dowling}},\ }\href {\doibase
  10.1103/PhysRevLett.104.103602} {\bibfield  {journal} {\bibinfo  {journal}
  {Phys. Rev. Lett.}\ }\textbf {\bibinfo {volume} {104}},\ \bibinfo {pages}
  {103602} (\bibinfo {year} {2010})}\BibitemShut {NoStop}%
\bibitem [{\citenamefont {Zhang}\ \emph {et~al.}(2013)\citenamefont {Zhang},
  \citenamefont {Jin}, \citenamefont {Cao}, \citenamefont {Liu},\ and\
  \citenamefont {Fan}}]{zhang13}%
  \BibitemOpen
  \bibfield  {author} {\bibinfo {author} {\bibfnamefont {Y.~R.}\ \bibnamefont
  {Zhang}}, \bibinfo {author} {\bibfnamefont {G.~R.}\ \bibnamefont {Jin}},
  \bibinfo {author} {\bibfnamefont {J.~P.}\ \bibnamefont {Cao}}, \bibinfo
  {author} {\bibfnamefont {W.~M.}\ \bibnamefont {Liu}}, \ and\ \bibinfo
  {author} {\bibfnamefont {H.}~\bibnamefont {Fan}},\ }\href
  {http://stacks.iop.org/1751-8121/46/i=3/a=035302} {\bibfield  {journal}
  {\bibinfo  {journal} {Journal of Physics A: Mathematical and Theoretical}\
  }\textbf {\bibinfo {volume} {46}},\ \bibinfo {pages} {035302} (\bibinfo
  {year} {2013})}\BibitemShut {NoStop}%
\bibitem [{\citenamefont {Tsang}(2012)}]{qzzb}%
  \BibitemOpen
  \bibfield  {author} {\bibinfo {author} {\bibfnamefont {M.}~\bibnamefont
  {Tsang}},\ }\href {\doibase 10.1103/PhysRevLett.108.230401} {\bibfield
  {journal} {\bibinfo  {journal} {Phys. Rev. Lett.}\ }\textbf {\bibinfo
  {volume} {108}},\ \bibinfo {pages} {230401} (\bibinfo {year}
  {2012})}\BibitemShut {NoStop}%
\bibitem [{\citenamefont {Giovannetti}\ \emph {et~al.}(2012)\citenamefont
  {Giovannetti}, \citenamefont {Lloyd},\ and\ \citenamefont
  {Maccone}}]{glm2012}%
  \BibitemOpen
  \bibfield  {author} {\bibinfo {author} {\bibfnamefont {V.}~\bibnamefont
  {Giovannetti}}, \bibinfo {author} {\bibfnamefont {S.}~\bibnamefont {Lloyd}},
  \ and\ \bibinfo {author} {\bibfnamefont {L.}~\bibnamefont {Maccone}},\
  }\href@noop {} {\bibfield  {journal} {\bibinfo  {journal} {Phys. Rev. Lett.}\
  }\textbf {\bibinfo {volume} {108}},\ \bibinfo {pages} {260405} (\bibinfo
  {year} {2012})}\BibitemShut {NoStop}%
\bibitem [{\citenamefont {Hall}\ \emph {et~al.}(2012)\citenamefont {Hall},
  \citenamefont {Berry}, \citenamefont {Zwierz},\ and\ \citenamefont
  {Wiseman}}]{hall2012}%
  \BibitemOpen
  \bibfield  {author} {\bibinfo {author} {\bibfnamefont {M.~J.~W.}\
  \bibnamefont {Hall}}, \bibinfo {author} {\bibfnamefont {D.~W.}\ \bibnamefont
  {Berry}}, \bibinfo {author} {\bibfnamefont {M.}~\bibnamefont {Zwierz}}, \
  and\ \bibinfo {author} {\bibfnamefont {H.~M.}\ \bibnamefont {Wiseman}},\
  }\href {\doibase 10.1103/PhysRevA.85.041802} {\bibfield  {journal} {\bibinfo
  {journal} {Phys. Rev. A}\ }\textbf {\bibinfo {volume} {85}},\ \bibinfo
  {pages} {041802} (\bibinfo {year} {2012})}\BibitemShut {NoStop}%
\bibitem [{\citenamefont {{Nair}}(2012)}]{nair2012}%
  \BibitemOpen
  \bibfield  {author} {\bibinfo {author} {\bibfnamefont {R.}~\bibnamefont
  {{Nair}}},\ }\href@noop {} {\bibfield  {journal} {\bibinfo  {journal} {ArXiv
  e-prints}\ } (\bibinfo {year} {2012})},\ \Eprint
  {http://arxiv.org/abs/1204.3761} {arXiv:1204.3761 [quant-ph]} \BibitemShut
  {NoStop}%
\bibitem [{\citenamefont {Giovannetti}\ and\ \citenamefont
  {Maccone}(2012)}]{gm_useless}%
  \BibitemOpen
  \bibfield  {author} {\bibinfo {author} {\bibfnamefont {V.}~\bibnamefont
  {Giovannetti}}\ and\ \bibinfo {author} {\bibfnamefont {L.}~\bibnamefont
  {Maccone}},\ }\href {\doibase 10.1103/PhysRevLett.108.210404} {\bibfield
  {journal} {\bibinfo  {journal} {Phys. Rev. Lett.}\ }\textbf {\bibinfo
  {volume} {108}},\ \bibinfo {pages} {210404} (\bibinfo {year}
  {2012})}\BibitemShut {NoStop}%
\bibitem [{\citenamefont {Knysh}\ \emph {et~al.}(2011)\citenamefont {Knysh},
  \citenamefont {Smelyanskiy},\ and\ \citenamefont {Durkin}}]{knysh}%
  \BibitemOpen
  \bibfield  {author} {\bibinfo {author} {\bibfnamefont {S.}~\bibnamefont
  {Knysh}}, \bibinfo {author} {\bibfnamefont {V.~N.}\ \bibnamefont
  {Smelyanskiy}}, \ and\ \bibinfo {author} {\bibfnamefont {G.~A.}\ \bibnamefont
  {Durkin}},\ }\href {\doibase 10.1103/PhysRevA.83.021804} {\bibfield
  {journal} {\bibinfo  {journal} {Phys. Rev. A}\ }\textbf {\bibinfo {volume}
  {83}},\ \bibinfo {pages} {021804} (\bibinfo {year} {2011})}\BibitemShut
  {NoStop}%
\bibitem [{\citenamefont {Escher}\ \emph {et~al.}(2011)\citenamefont {Escher},
  \citenamefont {de~Matos~Filho},\ and\ \citenamefont {Davidovich}}]{escher}%
  \BibitemOpen
  \bibfield  {author} {\bibinfo {author} {\bibfnamefont {B.~M.}\ \bibnamefont
  {Escher}}, \bibinfo {author} {\bibfnamefont {R.~L.}\ \bibnamefont
  {de~Matos~Filho}}, \ and\ \bibinfo {author} {\bibfnamefont {L.}~\bibnamefont
  {Davidovich}},\ }\href@noop {} {\bibfield  {journal} {\bibinfo  {journal}
  {Nature Physics}\ }\textbf {\bibinfo {volume} {7}},\ \bibinfo {pages} {406}
  (\bibinfo {year} {2011})}\BibitemShut {NoStop}%
\bibitem [{\citenamefont {{Escher}}\ \emph {et~al.}(2011)\citenamefont
  {{Escher}}, \citenamefont {{de Matos Filho}},\ and\ \citenamefont
  {{Davidovich}}}]{escher_bjp}%
  \BibitemOpen
  \bibfield  {author} {\bibinfo {author} {\bibfnamefont {B.~M.}\ \bibnamefont
  {{Escher}}}, \bibinfo {author} {\bibfnamefont {R.~L.}\ \bibnamefont {{de
  Matos Filho}}}, \ and\ \bibinfo {author} {\bibfnamefont {L.}~\bibnamefont
  {{Davidovich}}},\ }\href {\doibase 10.1007/s13538-011-0037-y} {\bibfield
  {journal} {\bibinfo  {journal} {Brazilian Journal of Physics}\ }\textbf
  {\bibinfo {volume} {41}},\ \bibinfo {pages} {229} (\bibinfo {year}
  {2011})}\BibitemShut {NoStop}%
\bibitem [{\citenamefont {Escher}\ \emph {et~al.}(2012)\citenamefont {Escher},
  \citenamefont {Davidovich}, \citenamefont {Zagury},\ and\ \citenamefont
  {de~Matos~Filho}}]{escher_prl}%
  \BibitemOpen
  \bibfield  {author} {\bibinfo {author} {\bibfnamefont {B.~M.}\ \bibnamefont
  {Escher}}, \bibinfo {author} {\bibfnamefont {L.}~\bibnamefont {Davidovich}},
  \bibinfo {author} {\bibfnamefont {N.}~\bibnamefont {Zagury}}, \ and\ \bibinfo
  {author} {\bibfnamefont {R.~L.}\ \bibnamefont {de~Matos~Filho}},\ }\href
  {\doibase 10.1103/PhysRevLett.109.190404} {\bibfield  {journal} {\bibinfo
  {journal} {Phys. Rev. Lett.}\ }\textbf {\bibinfo {volume} {109}},\ \bibinfo
  {pages} {190404} (\bibinfo {year} {2012})}\BibitemShut {NoStop}%
\bibitem [{\citenamefont {{Latune}}\ \emph {et~al.}(2012)\citenamefont
  {{Latune}}, \citenamefont {{Escher}}, \citenamefont {{de Matos Filho}},\ and\
  \citenamefont {{Davidovich}}}]{latune}%
  \BibitemOpen
  \bibfield  {author} {\bibinfo {author} {\bibfnamefont {C.~L.}\ \bibnamefont
  {{Latune}}}, \bibinfo {author} {\bibfnamefont {B.~M.}\ \bibnamefont
  {{Escher}}}, \bibinfo {author} {\bibfnamefont {R.~L.}\ \bibnamefont {{de
  Matos Filho}}}, \ and\ \bibinfo {author} {\bibfnamefont {L.}~\bibnamefont
  {{Davidovich}}},\ }\href@noop {} {\bibfield  {journal} {\bibinfo  {journal}
  {ArXiv e-prints}\ } (\bibinfo {year} {2012})},\ \Eprint
  {http://arxiv.org/abs/1210.3316} {arXiv:1210.3316 [quant-ph]} \BibitemShut
  {NoStop}%
\bibitem [{\citenamefont {{Demkowicz-Dobrza{\'n}ski}}\ \emph
  {et~al.}(2012)\citenamefont {{Demkowicz-Dobrza{\'n}ski}}, \citenamefont
  {{Ko{\l}ody{\'n}ski}},\ and\ \citenamefont {{Gu{\c t}{\u a}}}}]{demkowicz}%
  \BibitemOpen
  \bibfield  {author} {\bibinfo {author} {\bibfnamefont {R.}~\bibnamefont
  {{Demkowicz-Dobrza{\'n}ski}}}, \bibinfo {author} {\bibfnamefont
  {J.}~\bibnamefont {{Ko{\l}ody{\'n}ski}}}, \ and\ \bibinfo {author}
  {\bibfnamefont {M.}~\bibnamefont {{Gu{\c t}{\u a}}}},\ }\href {\doibase
  10.1038/ncomms2067} {\bibfield  {journal} {\bibinfo  {journal} {Nature
  Communications}\ }\textbf {\bibinfo {volume} {3}},\ \bibinfo {pages} {1063}
  (\bibinfo {year} {2012})},\ \Eprint {http://arxiv.org/abs/1201.3940}
  {arXiv:1201.3940 [quant-ph]} \BibitemShut {NoStop}%
\bibitem [{\citenamefont {{Knysh}}\ \emph {et~al.}(2014)\citenamefont
  {{Knysh}}, \citenamefont {{Chen}},\ and\ \citenamefont {{Durkin}}}]{knysh14}%
  \BibitemOpen
  \bibfield  {author} {\bibinfo {author} {\bibfnamefont {S.~I.}\ \bibnamefont
  {{Knysh}}}, \bibinfo {author} {\bibfnamefont {E.~H.}\ \bibnamefont {{Chen}}},
  \ and\ \bibinfo {author} {\bibfnamefont {G.~A.}\ \bibnamefont {{Durkin}}},\
  }\href@noop {} {\bibfield  {journal} {\bibinfo  {journal} {ArXiv e-prints}\ }
  (\bibinfo {year} {2014})},\ \Eprint {http://arxiv.org/abs/1402.0495}
  {arXiv:1402.0495 [quant-ph]} \BibitemShut {NoStop}%
\bibitem [{\citenamefont {Berry}\ \emph {et~al.}(2013)\citenamefont {Berry},
  \citenamefont {Hall},\ and\ \citenamefont {Wiseman}}]{bhw2013}%
  \BibitemOpen
  \bibfield  {author} {\bibinfo {author} {\bibfnamefont {D.~W.}\ \bibnamefont
  {Berry}}, \bibinfo {author} {\bibfnamefont {M.~J.~W.}\ \bibnamefont {Hall}},
  \ and\ \bibinfo {author} {\bibfnamefont {H.~M.}\ \bibnamefont {Wiseman}},\
  }\href {\doibase 10.1103/PhysRevLett.111.113601} {\bibfield  {journal}
  {\bibinfo  {journal} {Phys. Rev. Lett.}\ }\textbf {\bibinfo {volume} {111}},\
  \bibinfo {pages} {113601} (\bibinfo {year} {2013})}\BibitemShut {NoStop}%
\bibitem [{\citenamefont {Humphreys}\ \emph {et~al.}(2013)\citenamefont
  {Humphreys}, \citenamefont {Barbieri}, \citenamefont {Datta},\ and\
  \citenamefont {Walmsley}}]{humphreys}%
  \BibitemOpen
  \bibfield  {author} {\bibinfo {author} {\bibfnamefont {P.~C.}\ \bibnamefont
  {Humphreys}}, \bibinfo {author} {\bibfnamefont {M.}~\bibnamefont {Barbieri}},
  \bibinfo {author} {\bibfnamefont {A.}~\bibnamefont {Datta}}, \ and\ \bibinfo
  {author} {\bibfnamefont {I.~A.}\ \bibnamefont {Walmsley}},\ }\href {\doibase
  10.1103/PhysRevLett.111.070403} {\bibfield  {journal} {\bibinfo  {journal}
  {Phys. Rev. Lett.}\ }\textbf {\bibinfo {volume} {111}},\ \bibinfo {pages}
  {070403} (\bibinfo {year} {2013})}\BibitemShut {NoStop}%
\bibitem [{\citenamefont {Yuen}\ and\ \citenamefont {Lax}(1973)}]{yuen_lax}%
  \BibitemOpen
  \bibfield  {author} {\bibinfo {author} {\bibfnamefont {H.~P.}\ \bibnamefont
  {Yuen}}\ and\ \bibinfo {author} {\bibfnamefont {M.}~\bibnamefont {Lax}},\
  }\href {\doibase 10.1109/TIT.1973.1055103} {\bibfield  {journal} {\bibinfo
  {journal} {Information Theory, IEEE Transactions on}\ }\textbf {\bibinfo
  {volume} {19}},\ \bibinfo {pages} {740} (\bibinfo {year} {1973})}\BibitemShut
  {NoStop}%
\bibitem [{\citenamefont {Helstrom}\ and\ \citenamefont
  {Kennedy}(1974)}]{helstrom_kennedy}%
  \BibitemOpen
  \bibfield  {author} {\bibinfo {author} {\bibfnamefont {C.}~\bibnamefont
  {Helstrom}}\ and\ \bibinfo {author} {\bibfnamefont {R.}~\bibnamefont
  {Kennedy}},\ }\href {\doibase 10.1109/TIT.1974.1055173} {\bibfield  {journal}
  {\bibinfo  {journal} {Information Theory, IEEE Transactions on}\ }\textbf
  {\bibinfo {volume} {20}},\ \bibinfo {pages} {16} (\bibinfo {year}
  {1974})}\BibitemShut {NoStop}%
\bibitem [{\citenamefont {Genoni}\ \emph {et~al.}(2013)\citenamefont {Genoni},
  \citenamefont {Paris}, \citenamefont {Adesso}, \citenamefont {Nha},
  \citenamefont {Knight},\ and\ \citenamefont {Kim}}]{genoni}%
  \BibitemOpen
  \bibfield  {author} {\bibinfo {author} {\bibfnamefont {M.~G.}\ \bibnamefont
  {Genoni}}, \bibinfo {author} {\bibfnamefont {M.~G.~A.}\ \bibnamefont
  {Paris}}, \bibinfo {author} {\bibfnamefont {G.}~\bibnamefont {Adesso}},
  \bibinfo {author} {\bibfnamefont {H.}~\bibnamefont {Nha}}, \bibinfo {author}
  {\bibfnamefont {P.~L.}\ \bibnamefont {Knight}}, \ and\ \bibinfo {author}
  {\bibfnamefont {M.~S.}\ \bibnamefont {Kim}},\ }\href {\doibase
  10.1103/PhysRevA.87.012107} {\bibfield  {journal} {\bibinfo  {journal} {Phys.
  Rev. A}\ }\textbf {\bibinfo {volume} {87}},\ \bibinfo {pages} {012107}
  (\bibinfo {year} {2013})}\BibitemShut {NoStop}%
\bibitem [{\citenamefont {{Zhang}}\ and\ \citenamefont
  {{Fan}}(2014)}]{zhang14}%
  \BibitemOpen
  \bibfield  {author} {\bibinfo {author} {\bibfnamefont {Y.-R.}\ \bibnamefont
  {{Zhang}}}\ and\ \bibinfo {author} {\bibfnamefont {H.}~\bibnamefont
  {{Fan}}},\ }\href@noop {} {\bibfield  {journal} {\bibinfo  {journal} {ArXiv
  e-prints}\ } (\bibinfo {year} {2014})},\ \Eprint
  {http://arxiv.org/abs/1402.6197} {arXiv:1402.6197 [quant-ph]} \BibitemShut
  {NoStop}%
\bibitem [{\citenamefont {Van~Trees}\ and\ \citenamefont {Bell}(2007)}]{bell}%
  \BibitemOpen
  \bibinfo {editor} {\bibfnamefont {H.~L.}\ \bibnamefont {Van~Trees}}\ and\
  \bibinfo {editor} {\bibfnamefont {K.~L.}\ \bibnamefont {Bell}},\ eds.,\
  \href@noop {} {\emph {\bibinfo {title} {Bayesian Bounds for Parameter
  Estimation and Nonlinear Filtering/Tracking}}}\ (\bibinfo  {publisher}
  {Wiley-IEEE},\ \bibinfo {address} {Piscataway},\ \bibinfo {year}
  {2007})\BibitemShut {NoStop}%
\bibitem [{\citenamefont {Bell}\ \emph {et~al.}(1997)\citenamefont {Bell},
  \citenamefont {Steinberg}, \citenamefont {Ephraim},\ and\ \citenamefont
  {Van~Trees}}]{bell1997}%
  \BibitemOpen
  \bibfield  {author} {\bibinfo {author} {\bibfnamefont {K.}~\bibnamefont
  {Bell}}, \bibinfo {author} {\bibfnamefont {Y.}~\bibnamefont {Steinberg}},
  \bibinfo {author} {\bibfnamefont {Y.}~\bibnamefont {Ephraim}}, \ and\
  \bibinfo {author} {\bibfnamefont {H.}~\bibnamefont {Van~Trees}},\ }\href
  {\doibase 10.1109/18.556118} {\bibfield  {journal} {\bibinfo  {journal} {IEEE
  Transactions on Information Theory}\ }\textbf {\bibinfo {volume} {43}},\
  \bibinfo {pages} {624} (\bibinfo {year} {1997})}\BibitemShut {NoStop}%
\bibitem [{\citenamefont {Van~Trees}(2001)}]{vantrees}%
  \BibitemOpen
  \bibfield  {author} {\bibinfo {author} {\bibfnamefont {H.~L.}\ \bibnamefont
  {Van~Trees}},\ }\href@noop {} {\emph {\bibinfo {title} {Detection,
  Estimation, and Modulation Theory, Part I.}}}\ (\bibinfo  {publisher} {John
  Wiley \& Sons},\ \bibinfo {address} {New York},\ \bibinfo {year}
  {2001})\BibitemShut {NoStop}%
\bibitem [{\citenamefont {Fuchs}\ and\ \citenamefont {van~de
  Graaf}(1999)}]{fuchs}%
  \BibitemOpen
  \bibfield  {author} {\bibinfo {author} {\bibfnamefont {C.~A.}\ \bibnamefont
  {Fuchs}}\ and\ \bibinfo {author} {\bibfnamefont {J.}~\bibnamefont {van~de
  Graaf}},\ }\href@noop {} {\bibfield  {journal} {\bibinfo  {journal} {IEEE
  Trans. Inf. Theor.}\ }\textbf {\bibinfo {volume} {45}},\ \bibinfo {pages}
  {1216} (\bibinfo {year} {1999})}\BibitemShut {NoStop}%
\bibitem [{\citenamefont {Giovannetti}\ \emph {et~al.}(2003)\citenamefont
  {Giovannetti}, \citenamefont {Lloyd},\ and\ \citenamefont
  {Maccone}}]{glm_speed}%
  \BibitemOpen
  \bibfield  {author} {\bibinfo {author} {\bibfnamefont {V.}~\bibnamefont
  {Giovannetti}}, \bibinfo {author} {\bibfnamefont {S.}~\bibnamefont {Lloyd}},
  \ and\ \bibinfo {author} {\bibfnamefont {L.}~\bibnamefont {Maccone}},\ }\href
  {\doibase 10.1103/PhysRevA.67.052109} {\bibfield  {journal} {\bibinfo
  {journal} {Phys. Rev. A}\ }\textbf {\bibinfo {volume} {67}},\ \bibinfo
  {pages} {052109} (\bibinfo {year} {2003})}\BibitemShut {NoStop}%
\end{thebibliography}%
\end{document}